\documentclass{aa}  

\usepackage{graphicx}
\usepackage{txfonts}
\usepackage{lipsum}
\usepackage{subcaption}         
\usepackage{lscape}             
\usepackage{placeins}           

\usepackage{color}
\definecolor{gray}{RGB}{128,128,128}
\definecolor{darkred}{RGB}{128,0,0}
\newcommand{\gris}[1]{\textcolor{gray}{#1}}
\newcommand{\rougesombre}[1]{\textcolor{darkred}{#1}}

\newcommand{\G}{\mathcal{G}}

\newcommand{\Ms}{M_{\star}}

\newcommand{\Mp}{M_{\rm p}}
\newcommand{\Rp}{R_{\rm p}}
\newcommand{\dd}{\mathrm{d}}

\newcommand\Imag{\operatorname{Im}}

\usepackage{hyperref} 
\hypersetup{colorlinks=true,citecolor=blue}

\begin{document}

   \title{Estimation of the tidal heating in the TRAPPIST-1 planets}

   \subtitle{Influence of the internal structure}

   \author{Emeline Bolmont\inst{1,2}\fnmsep\thanks{Corresponding author: emeline.bolmont@unige.ch}
        \and Mariana Sastre\inst{1,2,3}
        \and Alexandre Revol\inst{1,2}
        \and Mathilde Kervazo\inst{1,2,4}
        \and Gabriel Tobie\inst{4}
        }

   \institute{Observatoire de Gen\`eve, Universit\'e de Gen\`eve, Chemin Pegasi 51, 1290, Sauverny, Switzerland 
   \and Centre pour la Vie dans l'Univers, Universit\'e de Gen\`eve, Geneva, Switzerland 
   \and Kapteyn Astronomical Institute, University of Groningen, P.O. Box 800, 9700 AV Groningen, The Netherlands
   \and Laboratoire de Plan\'etologie et G\'eosciences, UMR-CNRS 6112, Nantes Université, 44322 Nantes cedex 03, France}

   \date{Accepted in A\&A}

\abstract{
With the arrival of JWST observations of the TRAPPIST-1 planets—particularly secondary transit depths and phase curves—it is timely to reassess the contribution of tidal heating to their heat budget.
JWST thermal phase curves could reveal endogenic heating through an anomalously high nightside temperature, providing an opportunity to estimate tidal heating.

In this study, we revisit the tidal heating of these planets and propose a simple method to compute the tidal heating profile across a broad range of parameters.
Our approach leverages a known formulation for synchronously rotating planets on low-eccentricity orbits and the fact that the profile shape depends solely on internal structure.
We re-calculate the tidal heating contributions for all TRAPPIST-1 planets, with a particular focus on the impact of internal structure (core size and viscosity profile) and eccentricity uncertainties on their total heat budget.
Although the masses and radii of these planets are well constrained, degeneracies remain in their internal structure and composition.
For volatile-poor planets of silicate-rock compositions, we investigate the role of core iron content by exploring a range of core sizes.
For each structure, we compute the degree-two gravitational Love number, $k_2$, and the corresponding tidal heating profiles.
We assume sub-solidus temperatures profiles that are decoupled from interior heat production, which means our estimates are conservative and likely represent minimum values.

We find that the tidal heat flux for TRAPPIST-1 b and c can exceed Io’s heat flux, with uncertainties primarily driven by eccentricity.
These high fluxes may be detectable with JWST.
For planets f to g, the tidal flux remains below Earth’s geothermal flux, suggesting that tidal heating is unlikely to be the dominant energy source.
For planets d and e, however, tidal heating likely dominates their heat budget, potentially driving intense volcanic and tectonic activity, which could enhance their habitability prospects.
}

   \keywords{Planet-star interactions -- Planets and satellites: terrestrial planets -- Planets and satellites: interiors -- Planets and satellites: individual: TRAPPIST-1}

   \maketitle
\nolinenumbers

\section{Introduction}

Since the first detection of an exoplanet orbiting a solar-type star over two decades ago \citep{1995Natur.378..355M}, the number of detected exoplanets has increased rapidly, with several thousand now identified\footnote{\url{https://exoplanetarchive.ipac.caltech.edu/index.html}}. 
While initial efforts focused on exoplanet detection, the field is now shifting towards their characterization. 
Although the detection of small temperate exoplanets has been challenging, about 40 have been identified to date with some characteristics similar to Earth \citep[e.g.][]{2023A&A...679A.126T}, including those in the TRAPPIST-1 system \citep{2017Natur.542..456G}. 
This system provides an excellent opportunity to study the atmospheres and surfaces of rocky planets outside our solar system (e.g. with JWST, \citealt{2023Natur.618...39G,2023Natur.620..746Z}), which could revolutionize our understanding of the evolution and habitability of terrestrial planets through comparative planetology (e.g. \citealt{2020SSRv..216..100T}).

The TRAPPIST-1 (also referred to as T1) system consist{s} of seven Earth-sized planets orbiting a Jupiter-sized (very low mass) star forty light-years away \citep{2017Natur.542..456G}. 
At least three of these bodies orbit within the classical habitable zone of the star. 
These planets have semi-major axes ranging from 1 to 6\% of an astronomical unit, forming a compact and stable planetary system due to their resonant architecture \citep{2017NatAs...1E.129L}. 
An inherent particularity of the system due to its interacting planets is that, while transit photometry brought precise measurements of the planets’ sizes \citep{2018MNRAS.475.3577D,2020A&A...640A.112D}, their strong mutual interactions made possible to measure very precisely their {masses, hence giving access to their densities.}
Their composition is consistent with rocky compositions, more volatile-rich or iron-poor than the Earth’s, {at least for the outer planets} \citep{2018A&A...613A..68G,2018ApJ...865...20D,2021PSJ.....2....1A,2025Natur.638...69P}. 

Internal heating in rocky bodies shapes their interior and surface characteristics as well as their evolution. 
Among internal heating sources, tidal dissipation can represent a large source of energy for planetary interiors. 
The most striking evidence in the Solar System is the case of Io, archetype of tidally-heated world hosting extreme volcanism \citep[e.g.][]{2000Sci...288.1198S,de2019tidal,2021A&A...650A..72K}.
Interestingly, the orbital periods and eccentricities in the TRAPPIST-1 system are similar to those of the Galilean satellites, and both systems have mean-motion resonances. 
Tidal dissipation is therefore thought to contribute importantly to the total energy budget of the TRAPPIST-1 planets, with tidal heat fluxes being at least an order of magnitude larger than the Earth’s mean heat flux \citep{2018A&A...612A..86T,2020A&A...644A.165B}, and is expected to lead to the {persistence} of magma oceans in the closest TRAPPIST-1 planets \citep{2018A&A...613A..37B,2019A&A...624A...2D}. 
Given the recent claim of tidal heating occurring on the LP791-18d planet \citep{2023Natur.617..701P}, and the ongoing JWST observations of the TRAPPIST-1 system, it is now {timely} to reevaluate the estimation of the tidal heat flux for these planets.

This work thus generalizes to the work of \citet{2020A&A...644A.165B}, which was focused on the outer planets of TRAPPIST-1. 
Here, the use of the BurnMan code\footnote{available on \url{https://geodynamics.github.io/burnman/}} to compute the internal structure of the TRAPPIST-1 planets, allows us 1) to investigate the tidal response of the inner planets and 2) to investigate the impact of the degeneracy in the internal structure of the planets.
Furthermore, we use the up-to-date estimations of the radius, mass and eccentricities of the TRAPPIST-1 planets \citep{2021PSJ.....2....1A}.

We present our models, both internal structure model and tidal model in Section~\ref{Model} and give new estimates accounting for uncertainties in internal structure and eccentricities of the planet in Section~\ref{results_tidal_heating}.

\begin{table*}[h!]
\centering
    \caption{Core composition for {the TRAPPIST-1 planets}.}
    \label{tab:table1}
    \begin{tabular}{c|c|c|c|c|c|c|} 
      
Planet      & Excitation & Surface		& Smallest   & Earth-like core	  & Intermediate & Biggest 	 \\
            & frequency & Temperature		& core  & composition	&   &  core	 \\
		
      	&  (rad.s$^{-1}$)&	    & Fe, Si, S (\%)	   		& Fe, Si, S (\%)		& Fe, Si, S (\%)	 	& Fe, Si, S (\%)	 \\
      	\hline
      	\hline
T-1b      			&$4.804\times 10^{-5}$&300K & 92, 0, 0 	& 90, 2, 0	& 58, 18, 16	& 36, 29, 27  \\
R=1.116~R$_\oplus$	&$4.804\times 10^{-5}$&600K & 92, 0, 0	& 90, 2, 0	& 56, 19, 17 	& 36, 29, 27 \\
M=1.374~M$_\oplus$  &$4.804\times 10^{-5}$&670K & 92, 0, 0   & 90, 2, 0  & 58, 18, 16    & 36, 29, 27 \\
                	&$4.804\times 10^{-5}$&800K	& 92, 0, 0	& 90, 2, 0	& 58, 18, 16 	& 36, 29, 27  \\
	\hline
T-1c	      		&$2.999\times 10^{-5}$&300K	& 92, 0, 0	& 90, 2, 0	& 58, 18, 16 	& 36, 29, 27 \\
R=1.097~R$_\oplus$  &$2.999\times 10^{-5}$&600K	& 92, 0, 0	& 90, 2, 0	& 56, 19, 17    & 36, 29, 27 \\
M=1.308~R$_\oplus$	&$2.999\times 10^{-5}$&670K	& 92, 0, 0	& 90, 2, 0 	& 58, 18, 16	& 36, 29, 27 \\
                	&$2.999\times 10^{-5}$&800K	& 92, 0, 0	& 90, 2, 0 	& 58, 18, 16	& 36, 29, 27 \\
    	\hline
T-1d      			&$1.792\times 10^{-5}$&300K	& 92, 0, 0 	& 90, 2, 0  & 60, 17, 15 	& 36, 29, 27 \\
R=0.788~R$_\oplus$	&$1.792\times 10^{-5}$&600K	& 92, 0, 0 	& 90, 2, 0  & 60, 17, 15	& 36, 29, 27 \\
M=0.388~R$_\oplus$  &$1.792\times 10^{-5}$&650K	& 92, 0, 0	& 90, 2, 0  & 60, 17, 15	& 36, 29, 27 \\	
                    &$1.792\times 10^{-5}$&800K	& 92, 0, 0	& 90, 2, 0  & 60, 17, 15	& 36, 29, 27  \\	
	\hline
T-1e      			&$1.191\times 10^{-5}$&300K	& 92, 0, 0 	& 90, 2, 0  & 60, 17, 15    & 36, 29, 27 \\
R=0.920~R$_\oplus$	&$1.191\times 10^{-5}$&600K	& 92, 0, 0 	& 90, 2, 0 	& 60, 17, 15    & 36, 29, 27\\
M=0.692~R$_\oplus$  &$1.191\times 10^{-5}$&800K	& 92, 0, 0	& 90, 2, 0	& 60, 17, 15	& 36, 29, 27\\	
	\hline		
T-1f      			&$7.888\times 10^{-6}$&250K	& 92, 0, 0 	& 90, 2, 0 	& 58, 18, 16 	& 36, 29, 27 \\
R=1.045~R$_\oplus$	&$7.888\times 10^{-6}$&300K	& 92, 0, 0 	& 90, 2, 0 	& 58, 18, 16 	& 36, 29, 27 	\\
M=1.039~R$_\oplus$	&$7.888\times 10^{-6}$&600K	& 92, 0, 0 	& 90, 2, 0 	& 58, 18, 16	& 36, 29, 27   \\
                    &$7.888\times 10^{-6}$&800K	& 92, 0, 0	& 90, 2, 0	& 58, 18, 16	& 36, 29, 27 	\\	
	\hline
T-1g	      		&$5.878\times 10^{-6}$&210K	& 92, 0, 0 	& 90, 2, 0 	& 58, 18, 16	& 36, 29, 27 \\
R=1.129~R$_\oplus$	&$5.878\times 10^{-6}$&300K	& 92, 0, 0 	& 90, 2, 0 	& 58, 18, 16	& 36, 29, 27 \\
M=1.321~R$_\oplus$	&$5.878\times 10^{-6}$&600K	& 92, 0, 0 	& 90, 2, 0 	& 58, 18, 16	& 36, 29, 27  \\
                    &$5.878\times 10^{-6}$&800K	& 92, 0, 0	& 90, 2, 0	& 58, 18, 16	& 36, 29, 27 \\	
	\hline
T-1h      			&$3.869\times 10^{-6}$&170K	& 92, 0, 0 	& 90, 2, 0 	& 60, 17, 15    & 36, 29, 27 \\
R=0.755~R$_\oplus$  &$3.869\times 10^{-6}$&300K	& 92, 0, 0 	& 90, 2, 0 	& 60, 17, 15    & 36, 29, 27 \\
M=0.326~R$_\oplus$	&$3.869\times 10^{-6}$&600K	& 92, 0, 0 	& 90, 2, 0 	& 60, 17, 15    & 36, 29, 27 \\
                    &$3.869\times 10^{-6}$&800K	& 92, 0, 0	& 90, 2, 0	& 60, 17, 15    & 36, 29, 27 \\	
	\hline
	\end{tabular}
    \tablefoot{The radii and masses come from \citet{2021PSJ.....2....1A}. The percentage of iron, silicium and sulphur in the core is given.}
\end{table*}

\begin{figure*}[h]
\centerline{\includegraphics[width=\linewidth]{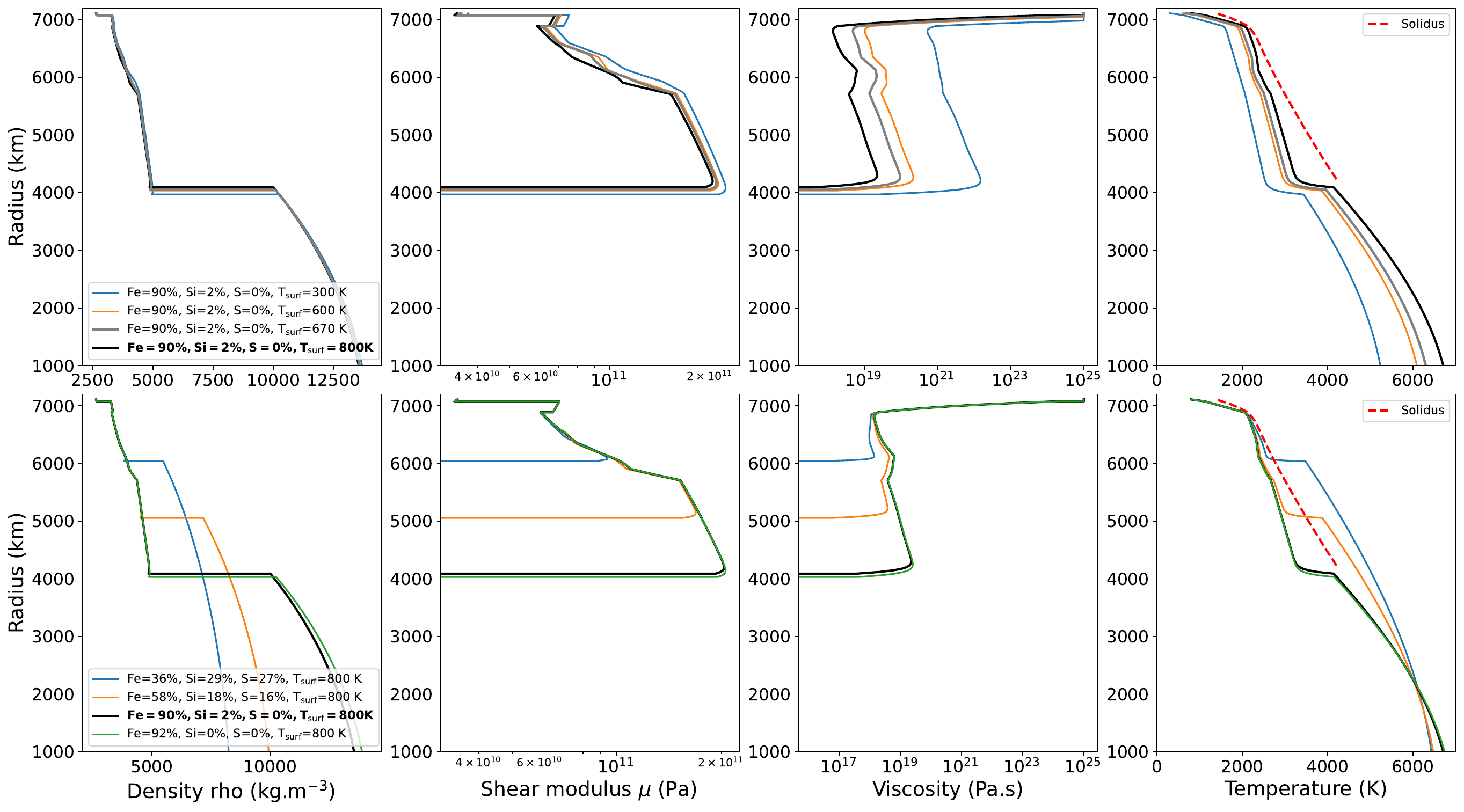}}
\caption{Density, shear modulus, viscosity and temperature profiles of TRAPPIST-1b computed with the BurnMan code for different compositions (listed in Table~\ref{tab:table1}). Top panel shows the influence of the temperature on the profile. Bottom panel shows the influence of the composition on the profile. The black curve shows the same composition in all panels (Earth-like composition and a surface temperature of 800~K). The gray curve in the top panels shows the curve corresponding to the reference temperature (here 670 K). The right most panels also show the temperature of the solidus (red dashed line).}
\label{Fig1}
\end{figure*}

\section{From internal structure to tidal heating}\label{Model}

In this section, we give the characteristics of both the internal structure model (Section~\ref{internal_struc}) we used and the model that allows us to derive (from the internal structure model) the gravitational Love number of degree 2 and thus the dissipation in the planet (Section~\ref{Love_Number}).
Finally, Section~\ref{tidal_heating} explains the method to compute the volumetric tidal heating inside the planets.

\subsection{Internal structure model}\label{internal_struc}

For a solid planet of given mass and radius, interior models estimate the mass fractions of its building materials - typically refractory elements like iron and silicates, as well as volatiles such as water and organics - distributed among a metallic core, a rocky mantle, and an external envelope of varying mass and thickness.
The composition of the planet has been proposed to be correlated to the metallicity of the host star \citep[e.g.][]{2018NatAs...2..297U}, but it might not be a perfect correlation with the composition of rocky planets spanning a wider distribution than that of stars \citep{2020MNRAS.499..932P}. 
In this work, we therefore adopt an agnostic strategy and consider a wide range of Fe/Si, resulting in a wide range of core sizes. 
As a result, in order to study a large range of solid exoplanet interior composition, several core sizes have been envisioned.

For each TRAPPIST-1 planet, we consider masses and radii from \citet{2021PSJ.....2....1A}. 
Despite the high precision of these estimations, internal structures and compositions are still degenerate and can be fit with various models.
In this study, we assume a silicate mantle with an Earth-like composition and a liquid metallic core characterized by a Fe, Si, S ratio.
To account for the degeneracy, we change the relative size of the core of the planets compared to the mantle, and we calculate their composition accordingly in order to reproduce both their mass and radius.
Our reference case is a planet that has the same core composition as the Earth, characterized by a Fe/Si mass ratio of 1.2 \citep[e.g.][]{1995ChGeo.120..223M}.
To keep the parameter space manageable, we do not investigate volatile-rich interior compositions, although these could also reproduce the planets’ masses and radii \citep{2021PSJ.....2....1A}.

We construct the internal structure models with the BurnMan code \citep{2014GGG....15.1164C, myhill_robert_2022_7080174}, which allows us to integrate consistently the physical properties such as density, temperature, gravity, and pressure from the surface to the planet's center, {assuming} hydrostatic equilibrium. 
This tool lets us compute self-consistently depth-dependent density, temperature, shear modulus and viscosity of each planet through a layered model consisting of, from the center to the surface, a liquid metallic core, a pyrolitic mantle and a dunite rigid crust.

BurnMan computes the 1D profiles of thermoelastic and thermodynamic properties with the Birch-Murnaghan finite-strain Equation of State (EoS). 
To do so, it uses lookup tables generated with Perple\_X \citep{2005E&PSL.236..524C} to access information on properties such as the heat capacity, conductivity, thermal expansivity, computed through thermodynamical equilibrium. 
BurnMan then uses this table to compute consistent profiles (density, thermal gradient, gravity gradient, P and S wave, among others) for different pressures and temperatures ranges.
For a few cases we computed, the pressure at the bottom of the mantle was higher than the maximum pressure of the built-in lookup table included in BurnMan (135~GPa). 
In that case, we expanded the lookup table using Perple\_X. 
While BurnMan can technically extrapolate EOS values beyond experimental calibrations, we caution that uncertainties increase at high pressures, particularly for silicates, due to the lack of experimental constraints at high pressure.
More details on the way the internal temperature profiles are calculated can be found in Appendix~\ref{App0}. 

Viscosity is a key parameter in determining tidal heating efficiency in planetary interiors \citep[e.g.][]{2020A&A...644A.165B}. 
In our model, viscosity is computed from a classical Arrhenius law, temperature dependent:
\begin{equation}\label{eq:viscosity_formula}
\eta = \frac{1}{2}A_0^{-1}d^{2.5}\exp{\left(\frac{E_a + PV_a}{RT}\right)},
\end{equation}
where $V_a$ is the activation volume, $E_a$ and $A_0$ are parameters depending on the material and $d$ is the grain size. 
Equation (1) is valid for diffusion creep. In this study, the viscosity prefactor and grain size are kept fixed for simplicity, although they are poorly constrained and could vary by orders of magnitude. 
Here, we consider dry olivine: $E_a = 300$~kJ.mol$^{-1}$ and $A_0 = 6.08 \times 10^{-19}$~Pa$^{-1}$.s$^{-1}$. 
We consider here a grain size d equal to 0.68~mm so that the factor $1/2~A_0^{-1}d^{2.5} = 10^{10}$~Pa.s.
For the activation volume, we follow the best fit solution pressure dependent proposed by \citet{2011Icar..216..572S}.
Using this prescription, typical values of $V_a$ span a range of 1.8~cm$^3$/mol (lower mantle) to 3.3~cm$^3$/mol (upper mantle) for T-1b.
As expected, this method leads to very high values of the viscosity near the surface, so we set the viscosity of the rigid crust to $10^{25}$~Pa.s. 
The viscosity of the inviscid liquid core is set to zero.

We first calculated a 1D ``Earth-like'' profile, which corresponds here to the internal structure the planets would have if they had the same composition as the Earth for the outer core  (``Earth-like'' column in Table~\ref{tab:table1}), mantle and lithosphere, from the center to the surface, respectively. 
The mantle is assumed to follow a perturbed adiabatic profile consistent with efficient internal cooling. This includes a basal Thermal Boundary Layer (TBL) with a contrast of 840 K. 
While plausible under Earth-like tectonics, such a TBL could be reduced in stagnant-lid planets or under strong tidal heating, which may suppress the gradient near the Core-Mantle Boundary (CMB). 
The liquid iron core is assumed to follow an adiabatic profile and extend from the planet's center to the core-mantle boundary. 
As previously said, we prescribed the mantle as a pyrolitic mantle. 
Finally, the lithosphere is assumed to have a dunite composition with a thickness of 200~km.

In the reference model, we assume a surface temperature of 300~K, a moho temperature (at the base of the crust) of 620~K and a temperature at the base of the lithosphere (lab) of 1550~K.
Then for each planet, we changed the amount of iron (Fe), silicon (Si) and sulfur (S) in the core, setting as a maximum amount of lighter elements, values constrained through chondritic models on Mercurian core \citep{2020JGRE..12506239V} and Mercurian models with eutectic composition \citep{2001Icar..151..118H}. 
In this way, we considered the smallest core possible (which therefore has the highest iron content possible to be able to reproduce the mass of the planet), the biggest core possible (which therefore has the lowest iron content possible to be able to reproduce the mass of the planet), an intermediate case and a Mercury-like case (where the core-size ratio is 85\%).
The extreme core case corresponds to a core size ratio even larger than the biggest core size measured in the solar system, which is why we also consider a Mercury-like case.
Note that for the Earth's core, Fe=90\%, Si=2\%, S=0\% \citep[e.g.][]{hirose2021light}.
For the different compositions, we vary Fe, Si and S while keeping the sum of Fe, Si and S at 92\%.
 
We also consider different surface temperatures for all planets: 300~K, 600~K, and 800~K, and a temperature which should be more representative of the surface temperature for each planet \citep[if they have an atmosphere {made of CO$_2$ or O$_2$ for instance, following} ][]{2018ApJ...867...76L}. 
These reference temperatures are: 670 K for TRAPPIST-1b (or T-1b) and c, 650 K for T-1d, 300 K for T-1e, 250 K for T-1f, 210 K for T-1g and 170 K for T-1h (M. Turbet, private communication, and compatible with \citealt{2018ApJ...867...76L}).
The increment in surface temperature is {then} passed on to {the moho and lab temperatures so that} if the surface temperature is increased by 300~K, so are the moho and lab temperatures. 
For the different possible surface temperatures, we calculate the profiles following the procedure described above for the different cases (smallest core, intermediate core, Mercury-like case, biggest core), ensuring that we always reproduce the observed radius and mass.
As the temperatures we considered are relatively similar, they have a small impact on the internal structure, so that the core compositions we obtain are sometimes very similar for different surface temperatures. 
This can be seen in Table~\ref{tab:table1}, where all the corresponding core compositions for the different core size fractions and surface temperatures can be found.  
The profiles of all planets assuming an Earth-like composition and {their corresponding reference surface temperatures} can be seen in Figure~\ref{FigA1}\footnote{All the other profiles can be found on \url{https://zenodo.org/records/14884378}.}. 

Figure~\ref{Fig1} shows the internal structure density profile for TRAPPIST-1b for the different temperatures and the core size assumptions.
The major differences can be seen for the different core sizes. 
In particular, when the core is small, both shear modulus and viscosity have a much wider range in the mantle. 
The increase in shear modulus is due to the increase of pressure with increasing mantle thickness, while the slight increase of viscosity is related to the temperature and pressure profile computed with Burnman. 
{As a result,} the viscosity is in average smaller if the planet has a big core than if it has a small core. 
Moreover, a bigger core and thinner mantle makes the mantle more deformable.

While variations in core size significantly impact both the shear modulus and viscosity, we find that changes in surface temperature have little impact on shear modulus but have a major impact on viscosity (top right panel).
Varying surface temperature thus allow to explore several viscosity distribution configuration, accordingly with Equation \ref{eq:viscosity_formula}.
Due to this temperature dependency, there are almost three order of magnitude between the viscosity we obtain at the base of the mantle between a surface temperature of 300~K and 800~K.
The different surface temperatures that we consider here, therefore, offer us a way to investigate the impact on different viscosity profiles on dissipation and the resulting tidal heating.

The right-most panels of Fig.~\ref{Fig1} display the silicate solidus temperature alongside the temperature profiles for TRAPPIST-1b (see App.~\ref{App1} for details and applications to the other planets). We find that the temperature profiles for TRAPPIST-1b remain below the solidus throughout the mantle for the range of surface temperatures explored. However, profiles with a surface temperature of 800~K approach the solidus at depth, suggesting that even a modest increase in internal heating could trigger partial melting.
While a full treatment of partial melting and its feedback on tidal dissipation is beyond the scope of this study, we note that approaching or crossing the solidus could substantially modify the planet’s tidal response. As shown by \citet{2021A&A...650A..72K}, the presence of partial melt can enhance tidal dissipation, particularly for melt fractions near the percolation threshold. 
Our present estimates focus on solid interiors and should therefore be regarded as lower limits of the actual tidal heat production.

\subsection{Calculating the frequency dependence of the Love number}\label{Love_Number}

To calculate the distribution of tidal dissipation in a spherical multilayer body, we use the same method here as in \citet{2005Icar..177..534T}, \citet{2020A&A...644A.165B} and \citet{2021A&A...650A..72K}. 
The method uses the elastic formulation of spheroidal oscillations developed by \citet{1972MetComPhy...1..217S}. 
Similar to \citep{2017JGRE..122.1338D}, we use the static formulation of \citep{Saito1974} for the liquid core. 
This method was adapted to the viscoelastic case by \citet{2005Icar..177..534T}, using the correspondence principle \citep{1954JAP....25.1385B}. 
It was recently used to study multilayer solid planetary interiors \citep{2019A&A...630A..70T,2020A&A...644A.165B, 2021A&A...650A..72K}.
We refer the reader to these publications, in particular Appendix A in \cite{2021A&A...650A..72K}, to have more details about the method.

We make the assumption that the viscoelastic response of the planets follows an Andrade rheology \citep{1910RSPSA..84....1A,2011JGRE..116.9008C}\footnote{Note that the Andrade rheology leads to higher values of the imaginary part of the Love number at the frequency range considered here (see Table~\ref{tab:table1}) compared to other rheologies like Maxwell \citep{2020A&A...644A.165B}. The difference is of several orders of magnitude which would then be passed on the total tidal heat budget.}. 
The Andrade rheological model is described by four parameters: the elastic shear modulus, $\mu$, the shear viscosity, $\eta$, and two additional parameters, $\alpha$, and $\beta$, describing the transient response between purely elastic and viscous response.  
The $\alpha$ parameter determines the frequency dependence of the viscoelastic response.
Following \citet{2019A&A...630A..70T} and \citet{2020A&A...644A.165B}, we assume a fiducial value of $\alpha$ of 0.25, which is a typical value reproducing the dissipation function of the Earth's mantle over a wide range of frequency. 
However, to account for the uncertainty on this parameter, we also consider values of $\alpha=0.20$ and $\alpha=0.30$.
For the $\beta$ parameter, following  \citep{2011JGRE..116.9008C}, we assume that $\beta \sim \mu^{\alpha-1} \eta^{-\alpha}$, which reproduces the existing data from mechanical tests on olivine minerals \citep{2001PCM....28..641T,2002JGRB..107.2360J}. 
Note, however, that more recent studies pointed out that significant uncertainties on the appropriate 
$\beta$ values remain, and $\beta$ may possibly vary between 0.01 and 100 $\times~\mu^{\alpha-1} \eta^{-\alpha}$ \citep{Bierson2024, Amorim2024}.

For each planet, {and the associated tested configurations in terms of core size fraction and surface temperature} (Table~\ref{tab:table1}), we calculate the dependence of the Love number with the excitation frequency. 
The excitation frequency $\omega$ is a linear combination of the mean motion of the planet $n$ and its spin $\Omega$. 
Here, we consider the rotation of the TRAPPIST-1 planets to be synchronous, their eccentricities small and their obliquity zero \citep[e.g.][]{2018A&A...612A..86T}. 
In that case, the main excitation frequency is $\omega = n$.
The main excitation frequency for each T1 planet is shown in Table~\ref{tab:table1}.

\begin{figure}[h]
\centerline{\includegraphics[width=\linewidth]{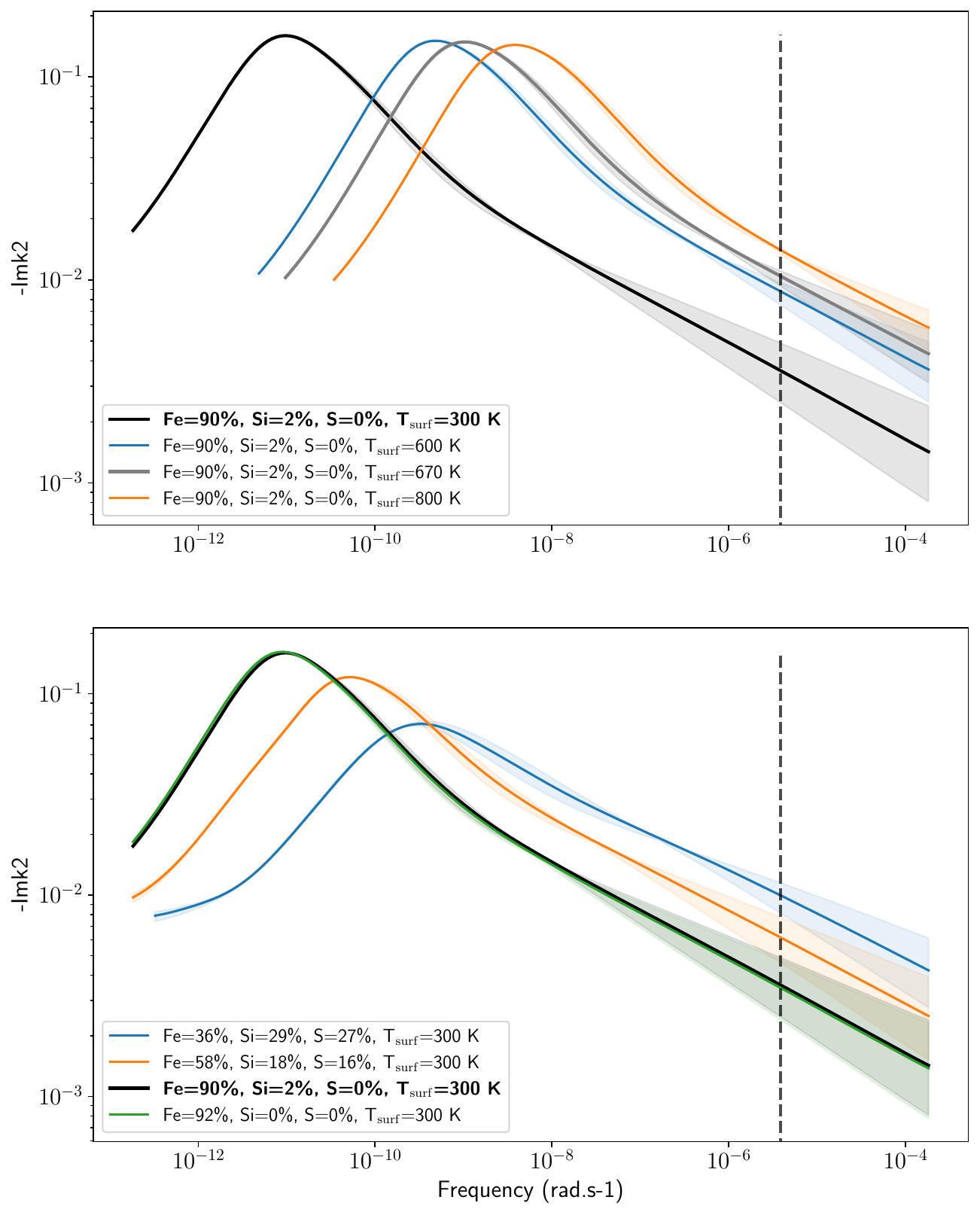}}
\caption{Frequency dependence of the imaginary part of the Love number for T-1b, for different compositions and temperatures (listed in Table~\ref{tab:table1}). Top panel shows the influence of the temperature. Bottom panel shows the influence of the composition. The black curve shows the same composition in both panels (Earth-like composition and a surface temperature of 300 K). The gray curve in the top panels shows the curve corresponding to the reference temperature (here 670 K). The excitation frequency of T-1b is shown as the vertical black dashed line, it corresponds to its orbital frequency. The shaded region illustrate the dependence of the imaginary part on $\alpha$ (bracketed between 0.20 and 0.30). At the frequency of the planet, the dissipation for $\alpha=0.20$ is higher than for $\alpha=0.30$.}
\label{Fig2}
\end{figure}

Figure~\ref{Fig2} shows the dependence of the imaginary part of the Love number Im(k$_2$) for T-1b.
As seen on the top panel, the impact of increasing the surface temperature (and thus decreasing the viscosity) has a strong impact on the Love number. 
As shown in \citet{2020A&A...644A.165B}, decreasing viscosity shifts the maximum of Im(k$_2$) to higher frequencies, which leads here to an increase of the value of Im(k$_2$) at the excitation frequency of the planet (shown as a vertical dashed black line).
In particular, between a surface temperature of 300~K and 800~K, Im(k$_2$) increases by about a factor 4 (for $\alpha=0.25$). 
The shaded areas in Fig~\ref{Fig2} show the dependence of the imaginary part of the Love number on $\alpha$. 
This parameter does not change the values of Im(k$_2$) for the lower frequencies but has an impact on the frequency range after the maximum of Im(k$_2$) \citep[see also Figure 2 from][]{2020A&A...644A.165B} where the slope is directly dependent on $\alpha$ (the slope is more pronounced for a high $\alpha$).
The differences grow the farther the frequency is from the frequency of the maximum. 
This means that for the highest surface temperature (orange curve), for which the maximum of Im(k$_2$) occurs closer to the excitation frequency, the impact of $\alpha$ on Im(k$_2$) is minimum.
Consistently, for the lowest surface temperature (black curve), for which the maximum of Im(k$_2$) occurs farther to the excitation frequency, the impact of $\alpha$ on Im(k$_2$) is maximum.

As seen on the bottom panel of Fig.~\ref{Fig2}, the impact of increasing the size of the core is to shift the frequency of the maximum dissipation to higher frequencies. 
This is expected as increasing the size of the core decreases the {average} viscosity of the mantle (see Fig.~\ref{Fig1}) and decreasing the viscosity has this effect \citep[see Figure 2 from][and top panel of this figure]{2020A&A...644A.165B}.
We can also see the impact of the presence and size of the core on the maximum of dissipation (around a frequency of $10^{-12}-10^{-11}$~rad.s$^{-1}$). 
This maximum increases for smaller cores.
This is also in agreement with the expected dependence of the imaginary part of the Love number with the shear modulus of the planet: the higher the shear modulus, the higher the maximum of the Love number \citep[see Figure 2 from][]{2020A&A...644A.165B}.
Indeed, Figure~\ref{Fig1} shows that the shear modulus is higher for smaller cores.
As for the top panel, due to the respective position of the peak compared to the excitation frequency, the impact of $\alpha$ on Im(k$_2$) is higher for the smallest core and lower for the biggest core.

As discussed earlier, Figure~\ref{Fig2} also shows the excitation frequency of T-1b as vertical black dashed lines. 
This frequency is higher than the frequency of the maximum of the imaginary part of the Love number, which is also the case for all the other planets including planet h which has the lowest excitation frequency. 
At that frequency, we can see that increasing the size of the core and the surface temperature increases the imaginary part of the Love number (bottom panel of Fig.\ref{Fig2}). 
Consequently, we would expect the highest tidal heating for the planets with the biggest core and the highest surface temperatures.
Note that for the two smaller cores cases (green and black curves), the dissipation at the frequency of T-1b is very similar, so we expect a similar tidal heating for these two internal structures.

 \subsection{Calculating the volumetric tidal heating}\label{tidal_heating}

We use the same method as in \citet{2020A&A...644A.165B} to compute the volumetric tidal heating $h_{\rm tide}$, which relies on Eq. 37 from \citet{2005Icar..177..534T} and is valid for synchronous, non-oblique planets on slightly eccentric orbits (${e\lesssim0.05}$)
\begin{equation}\label{htide}
h_{\rm tide}(r) = -\frac{21}{10} n^5 \frac{\Rp^4 e^2}{r^2} H_\mu \Imag \tilde\mu, 
\end{equation}
where $R_p$ is the radius of the planet, $r$ is the radius at which the volumetric tidal heating is estimated, $n$ is the orbital frequency and $e$ is the eccentricity of the orbit;
$H_\mu$ represents the radial sensitivity to the shear modulus $\mu$. 
{It depends on the radial structure of the planet and on $y_i$ functions, which are associated to radial and tangential displacements, radial and tangential stresses and the gravitational potential \citep{1972MetComPhy...1..217S}.
We refer the reader to \citet{2005Icar..177..534T,2020A&A...644A.165B} for a more in-depth explanation of the meaning of these $y_i$ functions. }
The $\Imag \tilde\mu$ in Eq. \ref{htide} is the imaginary part of the complex shear modulus. 
As in \citet{2005Icar..177..534T}, we assume that there is no bulk dissipation \citep[even if it might be an important effect in the case of partially molten interiors, see][]{2021A&A...650A..72K}, and that all dissipation is associated {with} shear deformation. 
In this formalism, $\Imag \tilde\mu$ contains all the information about the dissipation.
An important remark we can make here is that the radial dependency of the tidal heating is not dependent on the orbit and rotation of the planet, but on the internal structure only.

\begin{figure*}[h]
\centerline{\includegraphics[width=\linewidth]{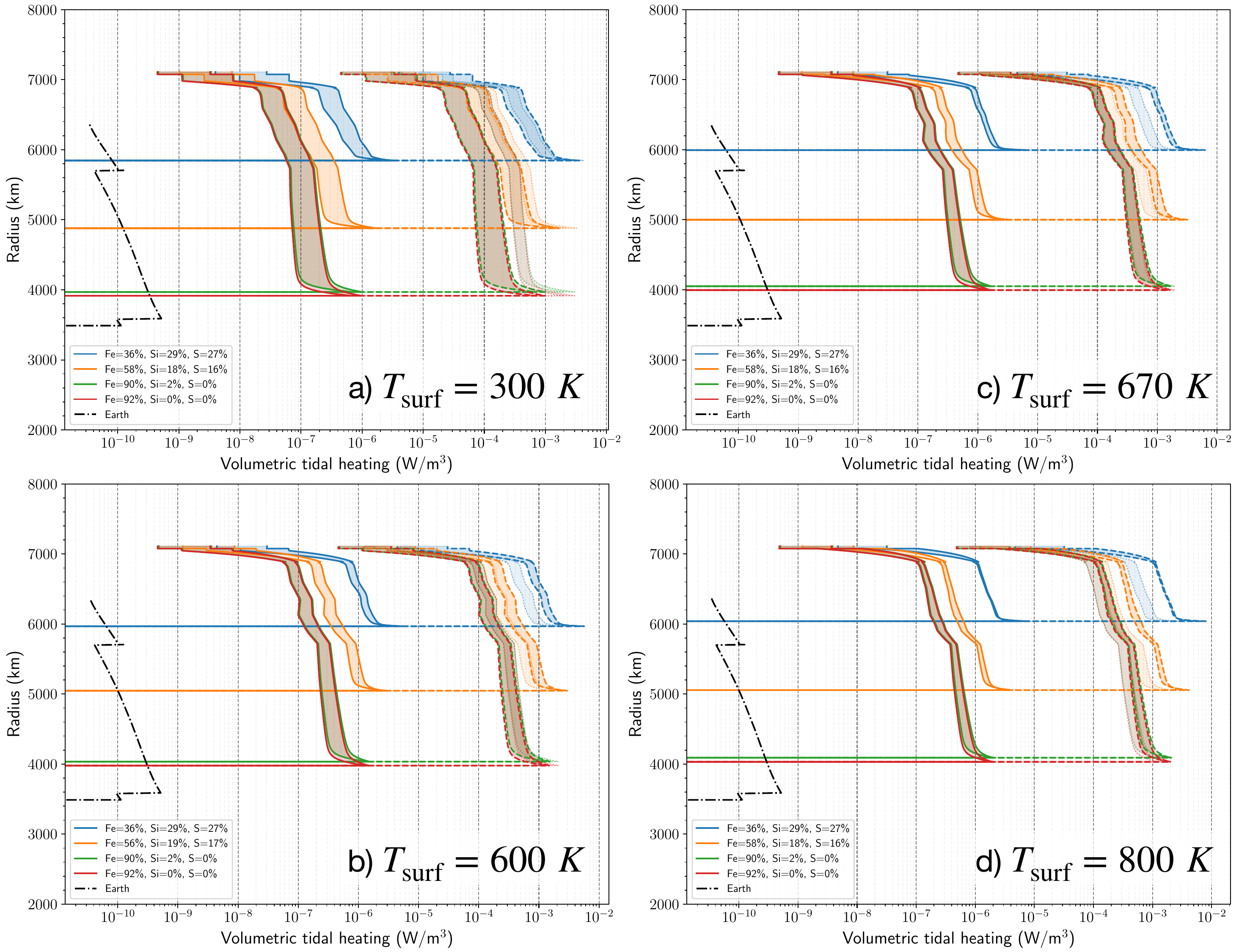}}
\caption{Volumetric tidal heating profile for T1-b for different surface temperatures (and thus viscosity profiles): a) 300~K, b) 600~K, c) 670~K, d) 800~K.
The different colors represent the different structures listed in Table~\ref{tab:table1}. 
The areas delimited by the full/dashed lines correspond to the minimum/maximum eccentricities given in Table~\ref{tab:table2}. 
The extent of the areas represents the sensitivity of the profile to $\alpha$, with the lower (left) limit corresponding to $\alpha=0.30$ and the upper (right) limit corresponding to $\alpha=0.20 $.
These profiles were obtained with Eq.~\ref{htide}. The tidal heating profile of the Earth is shown in a dashed black line as in \citet{2020A&A...644A.165B}.
Additionally, we represent areas delimited by faint dotted lines. 
These profiles are compatible with JWST observational constraints on the nightside temperature of the T1-b (291~K at 2$\sigma$, 322~K at 3$\sigma$ from \citealt{2025arXiv250902128G}), which are here hypothesized to be equal to a tidal temperature. The lower left limit thus corresponds to 291~K, and the upper right to 322~K.}
\label{Fig3}
\end{figure*}

Here we model only the dissipative power associated with the stellar eccentricity tides, and neglect other sources of dissipation due to obliquity tides, spin libration \citep{Frouard2017} or planet-planet tides \citep[][]{Hay2019}. 
Their contribution to the total dissipated power is expected to be small compared to the main eccentricity tides.
The obliquity of the planets has been shown to be very small \citep[$<1$ degree, see][]{2018A&A...612A..86T} and the amplitude of the spin librations were also estimated to be small \citep[$\lesssim1$ degree, see][]{2024A&A...691L...3R}. 
Previous studies suggest the possibility of large obliquity \citep{2024ApJ...961..203M} and large chaotic spin libration events \citep{2019MNRAS.488.5739V,shakespeare_day_2023,Chen_2023}.
However, the high-obliquity stable state found by \citet{2024ApJ...961..203M} requires unrealistically low dissipation for rocky material.
In addition, the results of \citet{2019MNRAS.488.5739V,shakespeare_day_2023,Chen_2023} rely on the CTL model, which is not well suited for studying the rotation of rocky planets \citep{2013ApJ...764...27M}.
A dedicated study would thus be necessary to make sure the heat from obliquity tides and spin librations are truly negligible, using a N-body code for instance \citep{2020A&A...635A.117B,2024A&A...691L...3R}.
Finally, the heat generated by planet-planet tides should be lower than $2.5\times10^{-2}$ times the contribution of eccentricity tides \citep{Hay2019}.
In any case, what we calculate here can be considered a lower estimate of the total heating, which should be slightly higher if we were to account for all these contributions. 

For the eccentricities, we consider two extreme cases coming from \citet{2021PSJ.....2....1A}. 
First, we consider what could be minimum eccentricities for the different planets, which are the forced eccentricities derived from the TTVs analysis performed in \citet{2021PSJ.....2....1A}.
These forced eccentricities can be considered as the eccentricities the planets would have if they had had time to be tidally damped.
Thus, these eccentricities solely arise from planet-planet interactions.
Second, we consider what could be maximum eccentricities for the different planets.
To compute these maximum eccentricities, we compute the quadratic sum of the forced eccentricities and the value of the free eccentricities for the 95th percentile (which corresponds to a 2$\sigma$ upper value).
The values of these eccentricities are given in Table~\ref{tab:table2}. 

To compute the total tidal heating, we can {either} integrate $h_{\rm tide}$ from Eq.~\ref{htide} over the planet as follows
\begin{equation}\label{Etide_int}
P_{\rm tide} = \iiint_V h_{\rm tide} \mathrm{d}V = 4\pi\int_{R_{\mathbf{CMB}}}^{R_p} h_{\rm tide}(r) r^2 \mathrm{d}r,
\end{equation}
where $R_{\mathbf{CMB}}$ is the radius of the core-mantle boundary{, or use a global formula as Eq.~2 of \citet{2005Icar..177..534T}}
\begin{equation}\label{Etide_Imk2}
P_{\rm tide} = -\frac{21}{2} \mathrm{Im}(k_2)\frac{(n R_{\mathrm p})^5}{G}e^2,
\end{equation}
The tidal heat flux is then $\Phi_{\mathrm{tide}} = P_{\rm tide}/(4\pi R_{\mathrm p}^2)$.
We have systematically checked the agreement of the global heat power between the two formulations of Eqs.~\ref{Etide_int} and \ref{Etide_Imk2}, and we get an error of less than 1.7\% between them.

\begin{table}[h!]
  \begin{center}
    \caption{Eccentricities considered to compute the tidal heating in the planets} 
    \label{tab:table2}
    \begin{tabular}{c|c|c|} 
Planet      & Minimum       & Maximum	 \\		
      	& eccentricity &eccentricity	 \\
      	& (forced) & (free)	 \\
      	\hline
T-1b      			&$2.77\times10^{-4}$  & $8.78\times10^{-3}$  \\
T-1c	      		&$5.23\times10^{-4}$  & $5.04\times10^{-3}$ \\
T-1d      			&$3.35\times10^{-3}$  & $5.54\times10^{-3}$ \\
T-1e      			&$6.88\times10^{-3}$  & $7.90\times10^{-3}$	\\
T-1f      			&$7.06\times10^{-3}$  & $7.66\times10^{-3}$	\\
T-1g	      		&$4.31\times10^{-3}$  & $4.91\times10^{-3}$ \\
T-1h      			&$2.18\times10^{-3}$  & $3.43\times10^{-3}$\\	
	\hline
	\end{tabular}
 \end{center}
\end{table}

\section{Tidal heating}\label{results_tidal_heating}

We first discuss our results in the framework of the previous section (synchronous rotation, no obliquity, and a small eccentricity), but in the second subsection, we propose a convenient way to calculate the heating profile for a generic case. 

\subsection{For synchronous rotation, no obliquity and a small eccentricity}\label{limited}

Figure~\ref{Fig3} illustrates the impact of 1) the four surface temperatures considered ($T_\mathrm{surf}$ = 300, 600, 670 and 800 K, from a to d panel respectively, 2) the various size of core of Table~\ref{tab:table1} (colored lines), 3) the different eccentricities of Table~\ref{tab:table2} (full and dashed lines) and 4) the $\alpha$ parameter (extent of the colored areas) on the heating profile inside T1-b\footnote{The Figures for all the other planets can be seen on \url{https://zenodo.org/records/14884378}.}.

Figure~\ref{Fig3} shows that increasing the size of the core increases the tidal heating in the mantle, with the maximum dissipation occurring at the base of the mantle, as observed in \citep[][]{2020A&A...644A.165B}.
We can see a difference of tidal heating at the base of the mantle of a factor 3-4 between the smallest core case (red) to the biggest core case (blue) for all surface temperatures.
This is compatible with Fig.~\ref{Fig2} in Section~\ref{Love_Number}, where the difference in Love number at the excitation frequency was about this order of magnitude.
Figure~\ref{Fig3} also shows the influence of the assumption on the eccentricity of the planet (full or dashed lines).
The maximum eccentricity of planet b is 32 times higher than the minimum eccentricity (see Table~\ref{tab:table2}) and this leads to a 3 order of magnitude difference in the tidal heating (as shows the $e^2$ factor in Eq.~\ref{htide}).
Finally, the extent of the colored areas between the full and dashed lines allows to visualize the impact of the $\alpha$ parameter, which corresponds to an uncertainty of a factor 2 in the tidal heating for a surface temperature of 300~K and the smallest core case. 
Consistently with what was discussed in Section~\ref{Love_Number}, this uncertainty due to $\alpha$ decreases for bigger cores.
As we were observing in Fig.~\ref{Fig2}, the impact of $\alpha$ decreases with increasing surface temperature.
While not negligible, especially for the lowest temperatures/higher viscosities, the uncertainty on $\alpha$ is lower than the uncertainty on the surface temperature/viscosity and on the internal structure and much lower than the uncertainty on the eccentricity.
While this is true for planet b, this might be different for the other planets for which the eccentricity is much better constrained than that of planet b.

\begin{figure*}
        \centering
   \includegraphics[width=\linewidth]{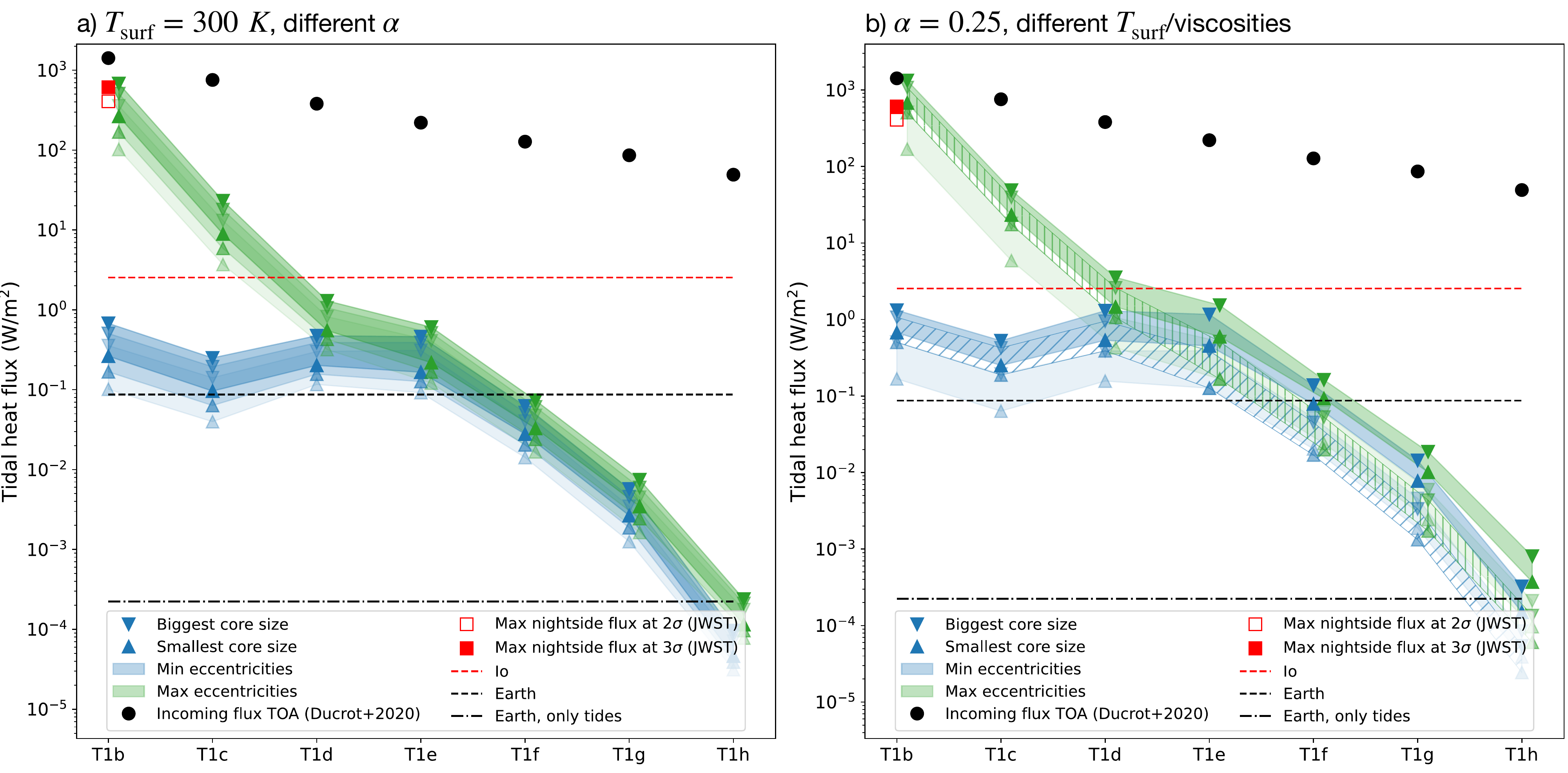}
      \caption{Total heat flux for all planets. Left panel: calculated for a surface temperature of 300~K (or a high viscosity). The transparency of the colored areas represents the dependency on $\alpha$, with the more transparent (lower values of tidal heating) corresponding to $\alpha=0.30$.
Right panel: calculated for an $\alpha=0.25$. The lighter shaded region (lower values of tidal heating) corresponds to a surface temperature of 300~K (high viscosities). The darker shaded region (higher values of tidal heating) corresponds to a surface temperature of 800~K (low viscosities). The hatched region corresponds to the reference temperatures (670~K for b and c, 650~K for d, 300~K for e, 250~K for f, 210~K for g and 170~K for h).
The colored area delimited by triangles represents the uncertainty we have on the internal structure, for a given assumption of the eccentricity (blue: minimum eccentricity, green: maximum eccentricity). These values are compared to the tidal heat flux of Io (full red triangle), Earth's heat flux (full black triangle), and the Earth's tidal heat flux (black triangle).
The Top Of the Atmosphere (TOA) fluxes coming from \citet{2020A&A...640A.112D} are shown as full black circles. 
Finally, we show as red squares recent observational constraints from the JWST \citep{2025arXiv250902128G} for the maximum nightside temperature of T1-b which we assume is a tidal temperature (see Section~\ref{generic}).}
\label{Fig5}
   \end{figure*}

A more convenient way to visualize the dependency of the tidal heating on the different parameters is to calculate the resulting total heat flux (calculated either by Eq.~\ref{Etide_int} or Eq.~\ref{Etide_Imk2}). 
Figure~\ref{Fig5} shows the tidal heat flux for all planets of TRAPPIST-1, calculated with Eq.~\ref{Etide_int}.
Once again, we represent the influence of the eccentricity (blue for low eccentricities and green for high) and of the internal structure (colored areas delimited by triangles)
Figure~\ref{Fig5}a shows the influence of the $\alpha$ parameter (different transparencies with $\alpha=0.30$ being the more transparent and $\alpha=0.20$ being the less transparent).
Figure~\ref{Fig5}b shows the influence of the surface temperature/viscosity profile (different transparencies with the lowest temperatures/highest viscosities being the more transparent and the highest temperatures/lowest viscosities being the less transparent).
The tidal heat flux is compared to the tidal heat flux of Io \citep[red dashed line, e.g.][]{2000Sci...288.1198S,2009Natur.459..957L}, the value of Earth's geothermal heat flux \citep[black dashed line][]{2010SolE....1....5D} and the Earth's tidal heat flux due to dissipation in the mantle (black dash-dotted line, computed from the profile in Fig.~\ref{Fig3}). 
Note that on the Earth, most of the tidal dissipation occurs in the ocean \citep[e.g.][]{Egbert2000,2003GeoRL..30.1907E}.
Values of the tidal heating of all planets can be found in Table~\ref{tab:table3}.
For this table, we give the values obtained for $\alpha=0.25$ and the most likely surface temperatures (670~K for b and c, 650~K for d, 300~K for e, 250~K for f, 210~K for g and 170~K for h) and the error bars encompass the rest of the uncertainties, meaning that the highest value proposed is the one obtained maximizing everything. 
So the maximum values correspond to the biggest core, $\alpha=0.10$ and the highest temperatures and the minimum values correspond to the smallest core, $\alpha=0.30$ and the lowest temperatures.
All combinations of parameters are given in the data accompanying this article\footnote{\url{https://zenodo.org/records/14884378}}.

Fig.~\ref{Fig5} shows that for planet b (and planet c), the minimum and maximum eccentricities are quite different (more than 1 order of magnitude), which leads to a huge difference in the tidal heat flux. 
The minimum values of the tidal heat flux are between the value for the heating of the Earth and the tidal heating of Io, which is the most volcanic body of our Solar System, while the maximum values are well above the tidal heat flux of Io.
Note that such high tidal fluxes are very likely to melt a large fraction of the rocky mantle, which would significantly impact the internal profile of the planet (both shear modulus and viscosity), and hence significantly modify the tidal response.  
In some circumstances, melt accumulation may result in the formation of mushy layers where tidal dissipation may be strongly enhanced {due to localized reductions in viscosity and shear modulus} \citep[e.g.][]{2021A&A...650A..72K} and possibly, in case of extreme melt production, to the formation of magma oceans where other mode of dissipation driven by gravito-inertial waves may develop, analogous to dissipation in the Earth's ocean \citep[e.g.][]{2003GeoRL..30.1907E, 2015ApJS..218...22T, 2022A&A...665L...1F,2024GeoRL..5107869A,2025ApJ...979..133F}. 
While a full treatment of partial or total melting and its dynamical feedbacks is beyond the scope of this study, we acknowledge its potential impact on the tidal response. 
In our models, temperature profiles are computed independently of the predicted tidal heating rates. 
However, for the most irradiated planets such as TRAPPIST-1b and c, the high volumetric tidal heating could drive temperatures toward or beyond the silicate solidus, leading to partial melting in the mantle. 
As shown in \citet[][]{2021A&A...650A..72K}, partial melt can enhance tidal dissipation, especially when the melt fraction is close to the critical melt fraction. 
Above this limit, however, the material may behave more fluid-like, reducing the efficiency of viscoelastic dissipation.

Fig.~\ref{Fig5} also shows that the uncertainty on the internal structure also leads here to a big uncertainty on the tidal heat flux for planets b and c, though in a lesser extent than the uncertainty on their eccentricity.
A factor of about 4-5 separates the flux of the smallest core case from the flux of the biggest core case.
For planets d, e, f, g and h, the uncertainty on the eccentricity is much smaller.
This means that for these planets, the uncertainty of the tidal flux is dominated by the uncertainty on the internal structure (i.e. the size of the core).

Let us focus on the impact of $\alpha$ (Fig.~\ref{Fig5}a) and the viscosity/surface temperature (Fig.~\ref{Fig5}b).
Figure~\ref{Fig5}a shows that the uncertainty on the $\alpha$ in the Andrade rheology leads to an uncertainty slightly lower than the uncertainty on the internal structure.
The difference of the global tidal heating between $\alpha=0.30$ and $\alpha=0.20$ is a factor 3-4, while the difference between the smallest core case and the biggest core case is of a factor 4-5. 
Figure~\ref{Fig5}b shows the impact of the viscosity profile chosen (via the surface temperature parameter).
The darker shaded area corresponds to the highest temperature for all planets considered here (800~K).
The light shaded area corresponds to a surface temperature of 300~K and the hatched area corresponds to the most likely surface temperature. 
The highest temperatures (lowest viscosities) lead to the highest tidal heat fluxes and the uncertainty this entails is of a factor 3-4 for the biggest core case (most dissipative structures) and a factor 4-6 for the smallest core case (least dissipative structure).

We can compare the tidal heat fluxes we calculate to the Top-Of-the-Atmosphere (TOA) fluxes for each planet from \citet{2020A&A...640A.112D}. 
These TOA fluxes are shown as full black circles in Figure~\ref{Fig5}.
For most planets, the TOA fluxes are much higher than the tidal heat flux.
However, assuming the highest eccentricity possible and the biggest core configuration for planet b and the highest surface temperature (the configuration which maximizes the tidal heat flux for a given $\alpha$, taken to be 0.25 for Fig.~\ref{Fig5}b), we obtain a tidal heat flux which is the same than the TOA flux. 
This means that a potential atmosphere of T1-b could be heated as much from the top than from the bottom, which should have repercussions on the atmosphere itself.
In any case, the tidal heat flux of planet b and to a lesser extent of planet c could be quite high and potentially observable on JWST emission or phase curve data of the system.

Figure~\ref{Fig5} shows that the tidal heat flux of T-1g and h should be lower than the heat flux of the Earth (black full triangle) and be as low as the tidal heat flux of the Earth (black triangle, note that it corresponds only to the rocky part, so that does not include the oceanic tide).
Taking into account other types of heating, like radioactive heating \citep[e.g.][]{2022ApJ...930L...6U}, induction heating \citep[e.g.][]{2017NatAs...1..878K}, or flare-induced heating \citep[e.g.][]{2022ApJ...941L...7G}, could potentially have a strong impact on the heat budget of the outer planets.
However, the inner planets should have a flux that is dominated by tidal heat flux (planets b to d).
Indeed, of the other sources of heating, it seems that flare-induced heating might be the most important in the context of TRAPPIST-1 and \citet{2022ApJ...941L..31V} showed that it is comparable to the energy released by Earth's radioactive elements today (so it should be comparable to the full black triangle value).

\renewcommand{\arraystretch}{1.5}
\begin{table*}[h!]
\centering
    \caption{Tidal heat flux (in W/m$^2$) for all the T1 planets.} 
    \label{tab:table3}
    \begin{tabular}{c|c|c|c|c|} 
Planet      &  \multicolumn{2}{c}{Minimum eccentricity}	                                        & \multicolumn{2}{c}{Maximum eccentricity}	 \\
            & Smallest core  & Biggest core                                                 & Smallest core  & Biggest core 	 \\
      	\hline
T-1b      	& ${0.50}^{+0.26}_{{-0.40}}$ 	& {$1.07^{+0.24}_{-0.71}$}   & $\rougesombre{505^{+266}_{-404}}$     & $\rougesombre{1074^{+245}_{-716}}$  \\
T-1c	    & $0.19^{+0.09}_{\gris{-0.15}}$	& {$0.43^{+0.08}_{-0.28}$}            & $\rougesombre{17.4}^{\rougesombre{+8.1}}_{\rougesombre{-13.7}}$    & $\rougesombre{39.5}^{\rougesombre{+7.1}}_{\rougesombre{-26.3}}$  	 \\
T-1d      	& $0.39^{+0.14}_{-0.27}$ 	& {$0.95^{+0.26}_{-0.65}$}                & $1.06^{+0.39}_{{-0.75}}$             & $\rougesombre{2.60}^{\rougesombre{+0.70}}_{-1.77}$  	 \\
T-1e      	& {$0.13^{+0.32}_{-0.03}$} 	& {$0.39^{+0.64}_{-0.08}$} 	              & $0.17^{+0.42}_{-0.05}$            & $0.51^{+0.85}_{-0.10}$ \\
T-1f      	& \gris{$0.017^{+0.062}_{-0.006}$} 	& $\gris{0.045}^{+0.082}_{\gris{-0.011}}$           	& $\gris{0.020}^{+0.073}_{\gris{-0.007}}$          & $\gris{0.053}^{+0.097}_{\gris{-0.013}}$  		\\
T-1g	    & \gris{$0.0013^{+0.0066}_{-0.0005}$} 	& \gris{$0.0033^{+0.0100}_{-0.0010}$}           & \gris{$0.0017^{+0.0086}_{-0.0006}$}      & \gris{$0.004^{+0.013}_{-0.001}$}  	 \\
T-1h      	& \gris{$0.000024^{+0.000115}_{-0.000006}$} 	& \gris{$0.000054^{+0.000237}_{-0.000013}$} & \gris{$0.00006^{+0.00029}_{-0.00002}$}      & \gris{$0.00014^{+0.00059}_{-0.00003}$}  \\
      	\hline

	\end{tabular}
    \tablefoot{The computations were done for $\alpha=0.25$ and the reference temperatures (670~K for b and c, 650~K for d, 300~K for e, 250~K for f, 210~K for g and 170~K for h). The uncertainties show the impact of $\alpha$ and the temperature/viscosity profile with the lower flux values corresponding to $\alpha=0.30$ and the lowest temperatures and the higher flux values corresponding to $\alpha=0.20$ and the highest temperatures. For comparison, the heat flux of Io is 2.54~W/m$^2$, the heat flux of the Earth is 8.7e-2~W/m$^2$ and the tidal heat flux of the Earth is 2.23e-4~W/m$^2$. Values in \gris{grey} identify where the flux is lower than the flux of the Earth, and values in \rougesombre{dark red} are higher than the flux of Io.}
\end{table*}

Assuming minimum eccentricities for the T1 planets, planets d and e have a similar or even higher flux than planet c. 
This is due to the fact that the forced eccentricity of planet c is about one order of magnitude lower than that of planet d and e (see Table~\ref{tab:table2}). 
This difference in eccentricity together with a higher value of the imaginary part of the Love number for their respective frequencies compensate the fact that planets d and e are farther from the star. 

Interestingly, T1-e, which might be the most apt to sustain an ocean of liquid water \citep[e.g.][]{2017ApJ...839L...1W,2018A&A...612A..86T,2020SSRv..216..100T}, has a tidal heat flux which could be of the order of magnitude of the heat flux of the Earth and higher.
This has strong implications on the possible habitability of the planet, as such heat fluxes could sustain volcanic activity and associated hydrothermal activities at the seafloor, similar to what has been shown on Jupiter's moon Europa \citep{2021GeoRL..4890077B}. 
It could also favor secondary outgassing, replenishing the atmosphere and surface in CO$_2$ and H$_2$O and thus favoring long-term habitability \citep[e.g.][]{2019A&A...625A..12G}.

\subsection{For non-synchronous rotation, non-zero obliquity and larger eccentricities}\label{generic}

Equations~\ref{htide} {and} \ref{Etide_Imk2} are valid for a specific set of orbital/rotational parameters. 
In particular, we used here expressions valid for a synchronous rotation, a zero obliquity, and small eccentricities \citep{2005Icar..177..534T}.
No expressions exist in the literature for a generic formula for the tidal heating profiles in multi-layered planets, probably due to the fact that it is extremely challenging to derive them. 
We therefore here propose a workaround for this limitation \citep[inspired from][]{2005Icar..177..534T}. 

The solution stems from the fact that the shape of the tidal heating profile only depends on the structure, while its amplitude of course depends on the excitation frequency and orbital/rotational parameters of the system. 
This means that if we have a way of calculating the global tidal heat flux of a planet for a generic case, it is possible to shift a profile previously computed under more restricted conditions so that its integral is equal to the global tidal heat flux. 
This allows us to assess a generic heating profile. 

This is what we have done in Figure~\ref{Fig3}. 
We have used observational constraints on the nightside temperature of TRAPPIST-1b obtained through the joint measurement of the phase curve of planets b and c \citep{2025arXiv250902128G}.
Given the amplitude of the reconstructed phase curve of planet b, and previous studies \citep{2023Natur.618...39G}, it seems very likely that it has no atmosphere. 
Whether the planet is tidally locked or not \citep{2019MNRAS.488.5739V}, the nightside should be very cold.
If it is tidally locked, it means the nightside is permanently in the dark. 
However, some studies have found that the T1 planets might not be perfectly tidally locked. 
For instance, \citet{2019MNRAS.488.5739V} proposed that their rotation is chaotic, while \citet{2024A&A...691L...3R} proposed that there is a slow drift of the substellar point, which therefore leads to a day-night cycle. 
However, arguments based on the thermal inertia of rocks and the proposed value for the sidereal day \citep[69 yr for T1-b, according to][]{2024A&A...691L...3R}, show that the day-night cycle is too long for the night-side to have retained some heat from its passage on the dayside.
We can therefore assume that a non zero temperature from the nightside would come from internal heating.
We consider here that the internal heating could be due to tides \citep[therefore neglecting other sources of heating such as induction heating,][]{2017NatAs...1..878K}.

The constraints in maximum nightside temperatures are $T_{2\sigma} = 291$~K at 2$\sigma$, $T_{3\sigma} = 322$~K at 3$\sigma$ from \citet{2025arXiv250902128G}. 
From these temperatures, we can obtain the corresponding global fluxes with $\Phi_{2\sigma} = \sigma_{\mathrm{SB}} T_{2\sigma}^4$ and $\Phi_{3\sigma} = \sigma_{\mathrm{SB}} T_{3\sigma}^4$, where $\sigma_{\mathrm{SB}}$ is the Stefan-Boltzmann constant.
Figure~\ref{Fig5} shows the values of these fluxes ($\Phi_{2\sigma} = 407$~W/m$^2$, $\Phi_{3\sigma} = 610$~W/m$^2$). 
They are relatively similar to the heat fluxes obtained previously for the maximum eccentricity.
However, we find some cases for which the tidal heat flux we obtain is higher than these observational constraints. 
This means that the combinations of structure/viscosity profile (surface temperature)/$\alpha$/eccentricity leading to higher flux values than $\Phi_{3\sigma}$ can probably be rejected.
If we consider that the maximum eccentricity allowed is the actual eccentricity of planet b (green color in Fig.~\ref{Fig5}), $\alpha=0.20$ and a high viscosity (300~K of surface temperature, Fig.~\ref{Fig5}a), we can reject the hot (low viscosity) biggest core profile (upper green triangle facing downwards).
If we consider that the maximum eccentricity allowed is the actual eccentricity of planet b, $\alpha=0.25$ (Fig.~\ref{Fig5}b), we can reject 1) all structures for the highest surface temperature/lower viscosities (darker shaded region), 2) all structures except for the smallest core case for a surface temperature of 650~K (hashed region).
Only the high viscosity structures are compatible with the observations, whatever the core size (lighter shaded region).

From these global flux values, we can then calculate the corresponding volumetric heat profile.
For each given internal structure, we start from a heat profile computed following the method described in the previous section~\ref{limited} (for instance, for the lowest eccentricity and $\alpha=0.25$), and we compute the corresponding global heat flux $\Phi_{\mathrm{ref}}$. 
We then compute the ratio of $\Phi_{2\sigma}$ and $\Phi_{3\sigma}$ to $\Phi_{\mathrm{ref}}$ and multiply the heat profile by this value.

Figure~\ref{Fig3} shows the resulting profiles for both observational constraints as the shaded area delimited by a semi-transparent dotted line: the left dotted curve corresponding to 291~K and right dotted curve corresponding to 322~K.
Once again, the different colors are for the different internal structures, and by construction these profiles all amount to the same global flux.

Here, we considered observational constraints for the global flux, but {the global flux} can also be computed from the Love number and the orbital/rotational parameters of the planet. 
The total energy can be written as the sum of the rotational energy and orbital energy, such as
\begin{equation}\label{eq:total_energy_spin+orbital}
    E_{\text{tot}} = E_{\text{rot}} + E_{\text{orb}} =\frac{1}{2} I \Omega_{\mathrm p}^2 -\frac{\G\Mp\Ms}{2a} ~.
\end{equation}

The loss of total energy is then evaluated with the time derivative as 
\begin{equation}\label{eq:tidal-heat-flux}
    \dot{E}_{\text{tot}} = I  \Omega_{\mathrm p} \frac{\dd \Omega_{\mathrm p}}{\dd t} + \frac{GM_sM_p}{2a^2}\frac{\dd a}{\dd t} = - \dot{E}_{\text{tide}} = -P_{\text{tide}}~,
\end{equation}
which corresponds to the amount of mechanical energy converted into tidal heat.
To be consistent with the way the profiles are here computed, the derivatives $\frac{\dd \Omega_{\mathrm p}}{\dd t}$ and $\frac{\dd a}{\dd t}$ can be obtained following \citet{2019CeMDA.131...30B}.
Using this way, one could compute the tidal heating profiles for any eccentricity, rotation or obliquity, given an internal structure, a value of $\alpha$ and a reference profile.

These steps are crucial for the next steps of this study which would be to take into account the impact of tidal heating on the interior structure of the planet and on its tidal dynamical evolution. 
Indeed, the tidal heating profile could be computed at each timestep of the integration of the orbit and rotation of the planet (for instance using ESPEM/SPIROID, e.g. \citealt{2023A&A...674A.227R}) and it could then be used to recompute a consistent interior structure. 
This interior structure could then be used to compute a new tidal Love number, which would allow us to compute the next timestep of the evolution of the semi-major axis, eccentricity, rotation, and obliquity of the planet.
This, however, is out of the scope of the present study.

\section{Conclusion}\label{sec:conclusion}

We here give new estimates of the tidal heat flux of the planets using the latest estimates of radii and masses \citep{2018MNRAS.475.3577D,2020A&A...640A.112D,2021PSJ.....2....1A}
and accounting for an Andrade rheology \citep{2020A&A...644A.165B}. 
We also propose a way to evaluate the tidal heating profile in planets which have parameters in rotation, eccentricity, obliquity which are not restricted to synchronous rotation, small eccentricities, and zero obliquity. 
The method we propose relies on the fact that the radial dependency of the profile only depends on the internal structure and the Andrade parameter $\alpha$ and that it is possible to compute a global heat flux in a generic way.
This is a first step to be able to one day study the retroaction of tidal heating on the internal structure of a multi-layered planet and on its corresponding tidal evolution. 
In particular, our models currently do not account for the thermomechanical feedback of tidal heating, such as temperature increase leading to partial melting. This omission may underestimate the actual heat flux in the most dissipative cases, especially for TRAPPIST-1b and c, where our predicted volumetric heating may locally approach or exceed the silicate solidus. 

Concerning the tidal heating estimates we provide here, we find that it could be higher than that of Io for planet b and c and that planets up to planet f could have tidal heat flux higher of the same order of magnitude of higher than Earth's heat flux. 
We also show that the uncertainty on the tidal heat flux is mainly due to the uncertainty on the eccentricity for planets b and c, and mainly due to the uncertainty on the internal structure (size of the core and viscosity profile) for the other planets. 
The uncertainties due to the internal structure are such that even with a very well observationally constrained system, the internal structure degeneracy is one major hurdle standing in the way of precise estimation of the tidal heating of planets.
Here we allowed for a wide range of different compositions \citep[though restricting ourselves to rocky compositions and therefore neglecting the fact that a volatile/water rich interior composition has been suggested,][]{2021PSJ.....2....1A}, but it might be possible to constrain this composition based on the composition of the star \citep[e.g.][]{2018ApJ...865...20D,2018NatAs...2..297U}. 
\citet{2021PSJ.....2....1A} actually proposes internal structures with Core Mass Fractions (CMF) of about 21$\pm$4 wt\% for all planets, which is close to the CMF we obtain for the Earth-case and the smallest core case.
However, even considering that the size of the core is known, the uncertainty on the viscosity profile (which we investigate via the surface temperature) is of a factor 4 (between the extremes we considered) and the uncertainty on the tidal parameter $\alpha$ is of a factor 2 (between the extremes we considered).

The existence of coreless terrestrial exoplanets has also been envisioned \citep[e.g.][]{2007ApJ...665.1413V,2007ApJ...659.1661F,2007ApJ...669.1279S,2007Icar..191..337S,2008ApJ...688..628E,2010A&A...516A..20V}. 
While no such configuration is observed among the planetary bodies of our Solar System, the discovery of exoplanets has broadened the range of possible interior compositions, motivating models of simple end-members. 
Our study, however, does not address this scenario. Indeed, 
in the case of the inner TRAPPIST planets, the hot interiors predicted by models would rule out this possibility, as these planets would have undergone melting early in their evolution, likely leading to the formation of a metallic core.
Additionally, \citet{2025arXiv251101231H} recently showed that oxygen partitioning rules out coreless TRAPPIST-1 planets. 

JWST observations (secondary transit depth or phase curve) of planet b and c could allow to bring constraints on the tidal heat flux of the planets, especially if those do not have an atmosphere or have very tenuous atmospheres, as might hint the first JWST observations of the inner planets \citep{2023Natur.618...39G,2023Natur.620..746Z}. 
Thanks to recent JWST observations \citep{2025arXiv250902128G}, we can reject low viscosity structures for T-1b if its eccentricity is high. 
Low viscosities lead to tidal heat fluxes higher than the constraints we have on the maximum nightside heat flux of the planet.
Additionally, TRAPPIST-1e experiences a tidal heat flux which could be compatible with volcanism or plate tectonics, which makes it an even more interesting astrobiological target. 
Future work should thus focus on self-consistent models that integrate tidal heating, interior melting, and thermal evolution, in order to better assess the long-term geodynamic and observational signatures of intense dissipation in close-in rocky exoplanets.

\section*{Data availability}

This work has made use of the BurnMan code, which is available on \url{https://geodynamics.github.io/burnman/}.
All interior profiles used in this study can be found on \url{https://zenodo.org/records/14884378}.

\begin{acknowledgements}
The authors would like to thank the anonymous referee for their constructive comments.
The authors would like to thank Lena Noack for the interesting discussions on viscosity, which led to some restructuring of the manuscript.
EB and AR acknowledge the financial support of the SNSF (grant number: 200021\_197176 and 200020\_215760).
This work has been carried out within the framework of the NCCR PlanetS supported by the Swiss National Science Foundation under grants 51NF40\_182901 and 51NF40\_205606.
MK and GT acknowledge the financial support of ERC PROMISES project (grant number: \#101054470).
The computations were performed at University of Geneva on the Baobab and Yggdrasil clusters.
This research has made use of the Astrophysics Data System, funded by NASA under Cooperative Agreement 80NSSC25M7105. 
\end{acknowledgements}

%

\begin{appendix}
\onecolumn

\section{More details about BurnMan computations}\label{App0}

The core is assumed to follow an adiabatic temperature gradient, consistent with a fully liquid metallic core. 
To compute this profile, we rely on BurnMan's built-in thermodynamic treatment of iron alloys. 
Specifically, we use the Fe–S Equation Of State (EOS) from \citet{2015JPCS...84...70S}, which is appropriate for planetary interior conditions up to several hundred GPa. 
To approximate the effect of light elements (e.g., S, Si), BurnMan applies a correction to the molar volume, which proportionally reduces the density and indirectly modifies thermodynamic properties such as the thermal expansivity and heat capacity. 
These quantities are then derived self-consistently from the EOS. 

For the mantle, the temperature profile is constructed by combining a perturbed adiabatic gradient in the convecting region with fixed thermal boundary layers at the surface and at the core-mantle boundary (CMB). 
The mantle itself is modeled using an Earth-like peridotitic composition, and its thermodynamic properties are obtained from BurnMan's implementation of the third-order Birch-Murnaghan EOS, based on Perple\_X-generated mineralogy and phase equilibria. 
The thermal profile assumes a total contrast of 900 K, with 60 K across the upper thermal boundary layer and 840 K across the basal boundary layer. 
This approach follows previous studies modeling Earth-like convecting planets and is consistent with a regime of vigorous mantle convection characterized by a Rayleigh number of $10^7$. 
Therefore, the CMB temperature and boundary layer thickness is not directly specified but results from the imposed thermal structure.

\section{Profiles for all planets assuming an Earth-like composition}\label{App1}

\begin{figure*}[ht!]
\centerline{\includegraphics[width=0.3\linewidth]{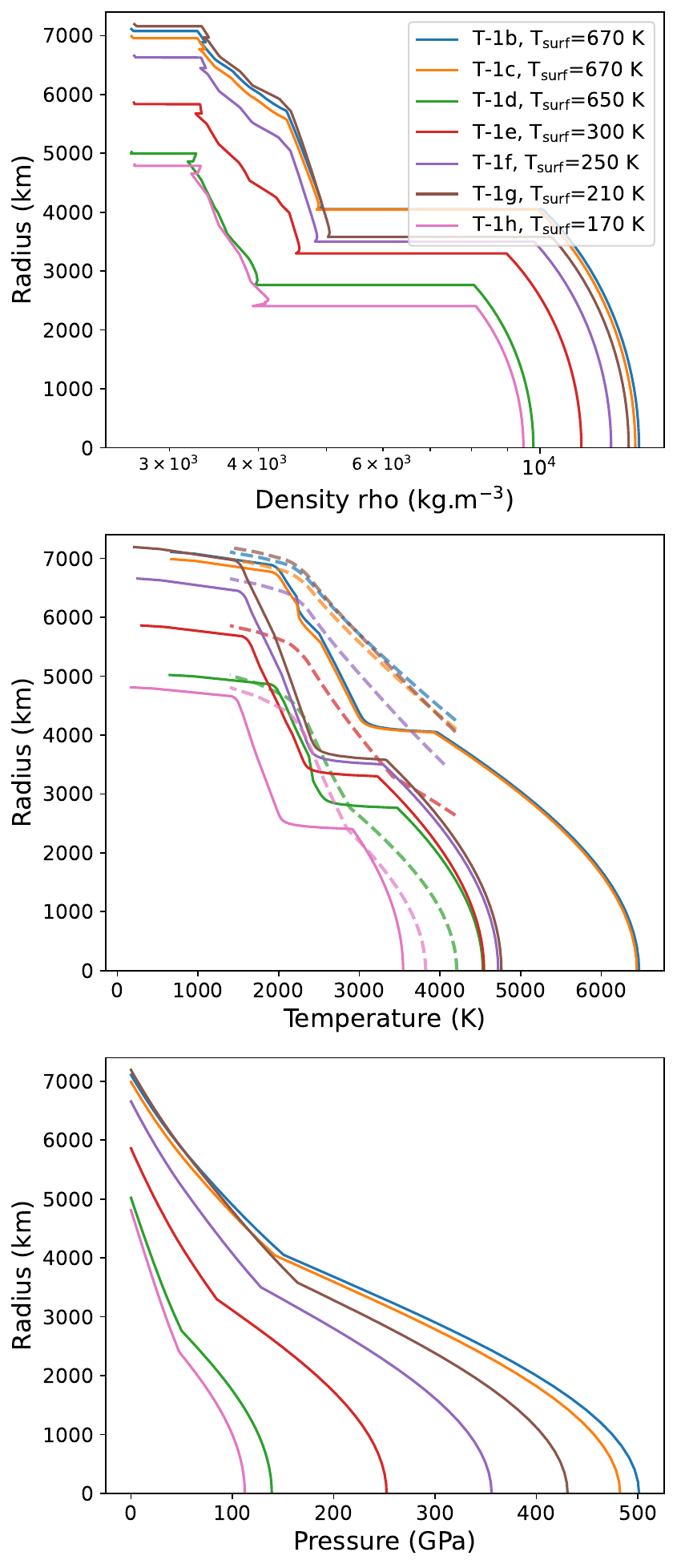}}
\caption{Profiles of density, temperature and pressure of the TRAPPIST-1 planets computed with the BurnMan code, assuming an Earth-like composition for the core and mantle. {The solidus is shown in dashed lines.}}
\label{FigA1}
\end{figure*}

Figure~\ref{FigA1} shows the different profiles of density, temperature and pressure of all the T1 planets assuming an Earth-like composition for the core.
We compare the temperature profiles computed with BurnMan to the silicate solidus, following \citet{2000GGG.....1.1042H} for pressures between 0 and 10~GPa, \citet{2000GGG.....1.1051H} between 10 and 22.5~GPa, and \citet{2016E&PSL.448..140M} from 22.5 to 136~GPa. For the surface temperatures considered here, all planets except TRAPPIST-1d exhibit mantle temperatures below the solidus, indicating no partial melting under the assumed conditions.
For TRAPPIST-1d, however, the temperature profile approaches or slightly exceeds the solidus near the base of the lithosphere, suggesting that limited partial melting could occur in this region.
Further work would be required to assess how such localized melting could affect the planet’s tidal response and internal heat transport \citep[e.g.][]{2021A&A...650A..72K,2021GeoRL..4890077B}. 

\section{Tidal temperature}\label{App3}

\begin{figure*}[ht!]
\centerline{\includegraphics[width=\linewidth]{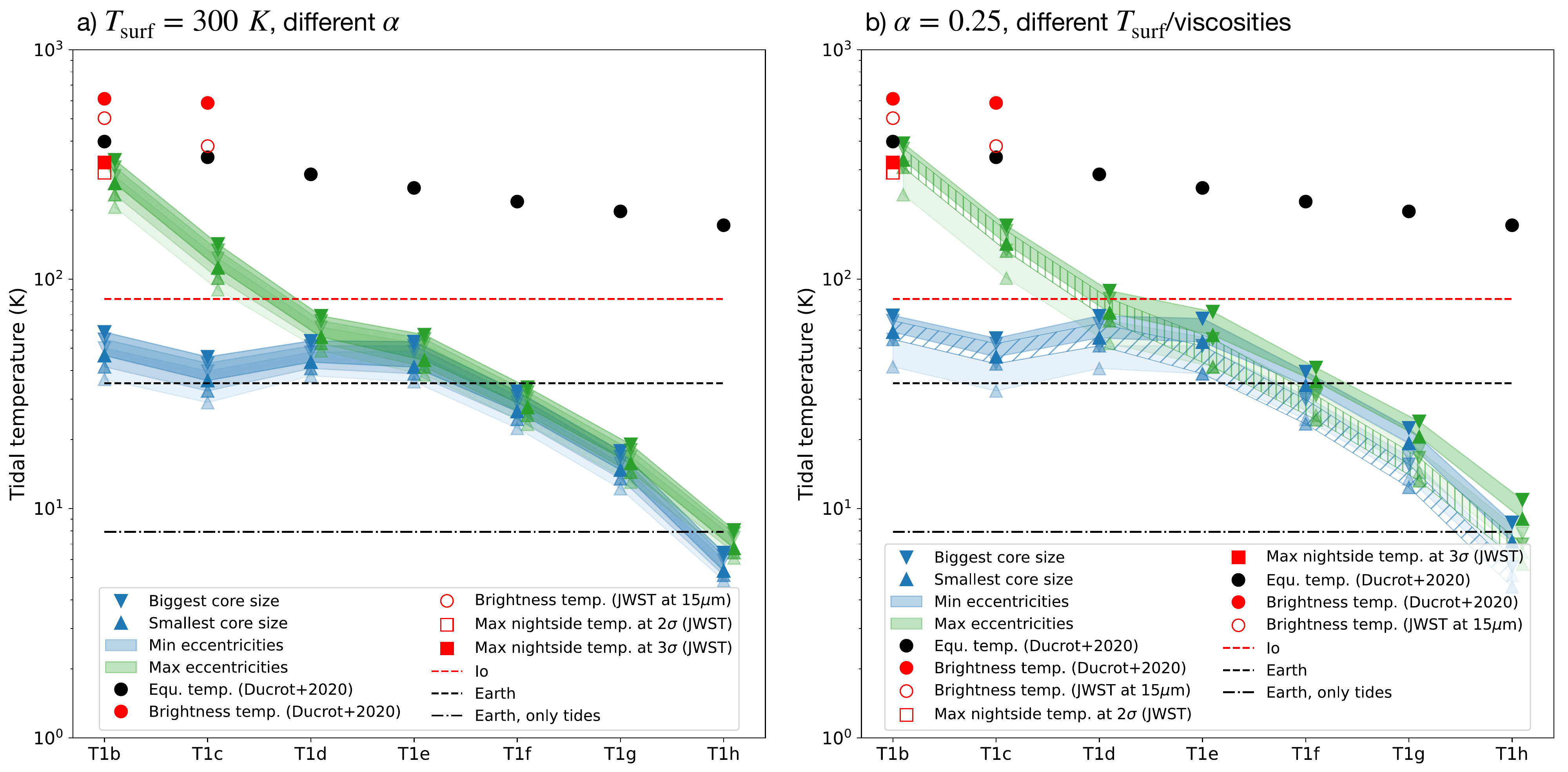}}
\caption{Same as Figure~\ref{Fig5} but for the tidal temperature. These values are compared with the equilibrium temperature of all planets (full black circles) and the brightness temperature of T-1b and c (full red circles) estimated in \citet{2020A&A...640A.112D}. We also give the recently acquired brightness temperatures measured by the JWST: the temperature of planet b comes from \citet{2023Natur.618...39G}, and that of planet c from \citet{2023Natur.620..746Z}.}
\label{FigA3}
\end{figure*}

Figure~\ref{FigA3} shows the tidal temperature for each planet, calculated as $\left(\Phi_{\mathrm{tide}}/\sigma_{\mathrm{SB}}\right)^{1/4}$, where $\Phi_{\mathrm{tide}}$ is the tidal heat flux of Fig~\ref{Fig5}. 
The tidal temperature is compared with the equilibrium temperature as estimated in \citet{2020A&A...640A.112D} and the brightness temperature for planets b and c \citep{2020A&A...640A.112D}.
We also compare for planet b the tidal temperatures we compute to the observational constraints on the maximum nightside temperature obtained by \citet{2025arXiv250902128G}.
The details can be found in the main text, Section~\ref{generic}.

Tides contribute to the temperature of the planets from a few Kelvin (for the outer planets) to more than a hundred Kelvin for the inner planets. 
Especially for T-1b, the maximum tidal temperatures are close to the maximum nightside temperature obtained from JWST constraints, and this can be translated into constraints on the degree of synchronization of the planets as well as on their obliquity \citep[see][]{2025arXiv250902128G}.

\FloatBarrier
\clearpage

\end{appendix}
\end{document}